# Experimental violation of a Bell-like inequality for causal order

**Yu Guo[1,2,3]†, Hao Tang[1,2,3]†, Bo-Xuan Wang[1,2,3], Min-Yu Lv[1,2,3], Jia-Wen Fan[1,2,3], Xiao-Min Hu[1,2,3,4], Yun-Feng Huang[1,2,3,4], Chuan-Feng Li[1,2,3,4], Guang-Can Guo[1,2,3,4], Giulio Chiribella[5,6,7,8]*, Bi-Heng Liu[1,2,3,4,9]***

**Quantum mechanics is compatible with scenarios where physical processes happen in an indefinite order. In theory, this feature could be detected through violations of inequalities on the observed correlations, analogous to Bell inequalities. However, experimental demonstrations of such violations have been missing until recently due to the complexity of the required setup. Here, we report an experimental violation of a Bell-like inequality involving the correlations of four parties, one of which is spacelike separated from the others. Our demonstration uses 3-kilometer fiber spools to simulate spacelike separation and achieves high-speed operations in photonic time-bin encoding, nanosecond synchronization, and accurate temperature stabilization. These experimental advances enable a violation by 5.7 SDs and open a path toward a certification of indefinite order in conditions that guarantee spacelike separation with existing state-of-the-art devices. However, the certification is not device independent, as it relies on knowledge about the setup to exclude bidirectional signaling—a loophole inherent to implementations in classical acyclic spacetimes, which may be resolved in future quantum-spacetime tests.**

## INTRODUCTION

Quantum mechanics is in principle compatible with scenarios in which two or more physical processes take place in an indefinite order (*1*–*4*). The prototype of this phenomenon arises when a quantum system is used to control the order in which two processes take place on a target system, giving rise to an operation known as the quantum switch (*2*, *4*). Over the past decade, the quantum switch has been the object of extensive research, both theoretical and experimental (*5*), which unveiled fundamental implications for spacetime physics (*6*–*9*), causal modeling (*10*), and time delocalization (*11*, *12*), and established a variety of advantages in quantum information tasks such as quantum channel discrimination (*13*), promise problems (*14*–*16*), communication complexity problems (*17*, *18*), quantum communication (*19*–*22*), quantum metrology (*23*–*25*), quantum thermodynamics (*26*–*35*), and others (*36*, *37*). Motivated by these applications, a series of works proposed methods for detecting indefinite order, including causal witnesses (*38*–*45*), process tomography (*46*), semidevice independent methods (*47*–*49*), and self-testing (*50*).

A natural question is whether the presence of indefinite order in the quantum switch could be experimentally detected through the violation of inequalities on the correlations between measurement outcomes observed in different laboratories, in a similar way as the presence of quantum nonlocality can be detected through the violation of Bell inequalities (*51*). A first analog of Bell inequalities for causal order was developed by Oreshkov *et al.* (*3*), who showed that the correlations between experiments performed in a well-defined order must obey nontrivial upper bounds called causal inequalities. However, the quantum switch does not violate any causal inequality (*38*, *52*), and therefore its indefinite causal order cannot be detected by this approach. A way around the problem was recently found in (*53*, *54*), which extended the framework of causal inequalities to new scenarios involving an additional spacelike separated party. In particular, van der Lugt, Barrett, and Chiribella (VBC) (*53*) introduced a Bell-like inequality, now known as the VBC inequality, that is maximally violated by the quantum switch. Using the VBC inequality, the indefinite causal order in the quantum switch can be detected directly from the correlations between measurement outcomes, potentially even in a device-independent way, using uncharacterized quantum devices.

Experimental violations of the VBC inequality, however, are challenging to achieve in the laboratory: They require a high-quality realization of the quantum switch, high-quality entanglement, and spacelike separation between the operations performed inside the quantum switch and those performed on another quantum system. In particular, the requirement of spacelike separation demands fast and coordinated measurement operations to be performed in different parts of the setup.

Here, we report an experimental violation of the VBC inequality in a photonic quantum switch, exceeding the VBC bound by 5.7 SDs while implementing fast operations inside the quantum switch. If combined with existing state-of-the-art detectors outside the quantum switch setup, such fast operations can guarantee spacelike separation of the additional party. Our setup performs measurements and state preparations of a time-bin degree of freedom (DoF) at a repetition rate of 0.1 MHz and integrates the synchronized operation of multiple components, including an acousto-optic modulator-based wave shaper, high-speed random number generators (RNGs), optical switching elements, and time-to-digital converters (TDCs).

[1]Laboratory of Quantum Information, University of Science and Technology of China, Hefei 230026, China. [2]CAS Center For Excellence in Quantum Information and Quantum Physics, University of Science and Technology of China, Hefei 230026, China. [3]Anhui Province Key Laboratory of Quantum Network, University of Science and Technology of China, Hefei 230026, China. [4]Hefei National Laboratory, University of Science and Technology of China, Hefei 230088, China. [5]QICI Quantum Information and Computation Initiative, School of Computing and Data Science, The University of Hong Kong, Pokfulam Road, Hong Kong. [6]Department of Computer Science, University of Oxford, Wolfson Building, Parks Road, Oxford, UK. [7]HKU-Oxford Joint Laboratory for Quantum Information and Computation, The University of Hong Kong, Pokfulam Road, Hong Kong. [8]Perimeter Institute for Theoretical Physics, 31 Caroline Street North, Waterloo, Ontario, Canada. [9]College of Physics, Guizhou University, Guiyang 550025, China.
*Corresponding author. Email: giulio@cs.hku.hk (G.C.); bhliu@ustc.edu.cn (B.-H.L.)
†These authors contributed equally to this work.

In addition, active temperature stabilization (with fluctuations maintained below 0.05°C) ensures high phase stability, enhancing the visibility of the interference between two causal orders. Last, 3-km fiber spools combined with 10-cm free-space delay lines simulate spacelike separation with the additional party involved in the VBC inequality by the distance traveled by the photons.

Overall, the experimental advances demonstrated in our work show that high-quality coherence, entanglement, and coordinated ultrafast operations can be combined to achieve a statistically notable violation of the VBC inequality. It is important to stress, however, that while VBC inequality violations could in principle enable a device-independent proof of indefinite causal order, the demonstration provided in our experiment is not device independent or loophole free as the latest generation of Bell nonlocality experiments (*55*–*57*). The main loophole here is intrinsic to the photonic realizations of the quantum switch, in which the laboratories of Alice 1 and Alice 2 must accept input photons for an extended period of time (*7*, *11*, *58*–*61*). During this time, Alice 1's settings could affect Alice 2's outcomes and vice versa, opening a loophole in which the violation of the VBC inequality arises from bidirectional signaling between the laboratories of the two Alices, rather than indefinite causal order of the operations performed by them. Whether a fully device-independent, loophole-free violation of the VBC inequality is experimentally achievable in nature remains as a major open question for future investigations.

## RESULTS

### Theoretical framework

Let us start by reviewing the settings of the VBC inequality. The inequality involves four parties, Alice 1, Alice 2, Bob, and Charlie, performing experiments with settings $x_1, x_2, y, z$, respectively, and outcomes $a_1, a_2, b, c$, respectively. In the derivation of the inequality, the duration of each experiment is assumed to be short enough that the experiment can be treated as localized in the neighborhood of a spacetime point. Specifically, VBC considered the situation in which Charlie's experiment takes place in the future lightcone of Alice 1 and Alice 2, while Bob is spacelike separated from the other three parties, as illustrated in Fig. 1A.

In the above settings, the VBC inequality follows from three basic assumptions: definite causal order (D), relativistic causality (R), and free interventions (F). D is the assumption that there exists a hidden variable λ specifying the causal order of the four parties. In particular, λ specifies whether Alice 1 acts in the causal past of Alice 2 or vice versa. R is the statement that the causal order of the four parties respects the light-cone structure in Fig. 1A. Last, F requires that the settings of the four parties have no relevant causes, and the probability distribution of the variables involved in the experiment satisfies the statistical independence relations implied by the causal structure. In conjunction with D, this assumption implies that the settings $x_1, x_2, y, z$ are independent of λ, and the conditional probability distribution $p(a_1, a_2, b, c \mid x_1, x_2, y, z, \lambda)$ allows for signaling from a setting $s$ to an outcome $o$ only if $s$ causally precedes $o$ in the causal order specified by λ.

VBC showed that the above assumptions imply a set of nontrivial inequalities, called DRF inequalities (*53*). On the other hand, quantum processes with indefinite causal order can violate (D) and therefore can violate DRF inequalities while still satisfying the standard

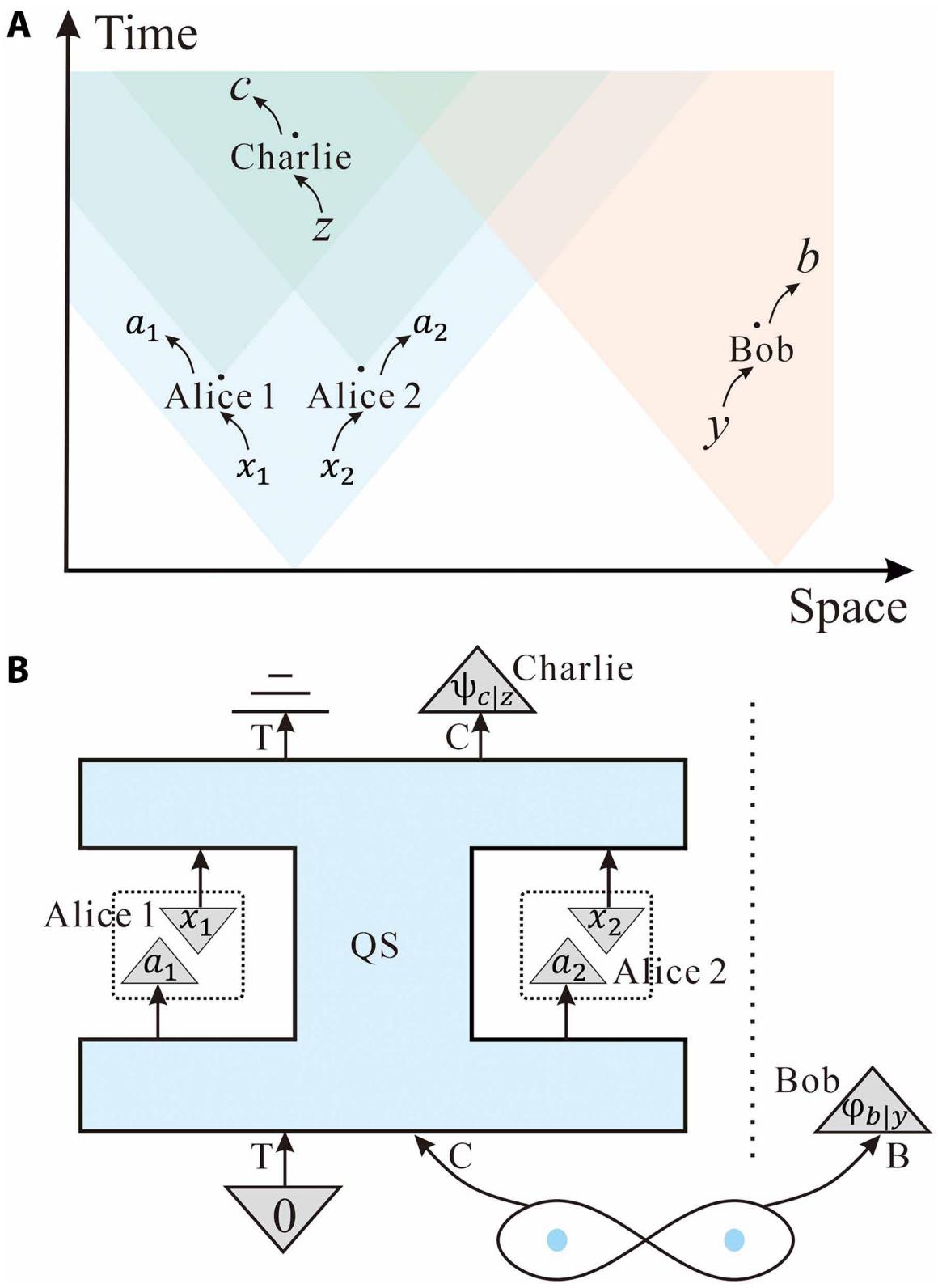

**Fig. 1. Light-cone structure and quantum switch setup for the VBC scenario.** (**A**) Light cone structure: Alice 1 and Alice 2 are the causal past of Charlie, while Bob is spacelike separated from the other three parties. (**B**) Quantum switch setup: A control qubit (C), initially maximally entangled with a distant qubit (B), determines the order in which a target system (T) undergoes the operations of Alice 1 and Alice 2. The target qubit starts in the state $|0\rangle$. The operations of the two Alices consists of projective measurements on the canonical basis $\{|0\rangle, |1\rangle\}$ (in which Alices 1 and 2 obtain outcomes $a_1$ and $a_2$, respectively) followed by state repreparations (in which Alice 1 and Alice 2 reset the target system to the states $|x_1\rangle$ and $|x_2\rangle$, respectively). Bob and Charlie perform the optimal quantum measurements for the violations of the CHSH inequality on systems B and C, with settings $y$ and $z$, and outcomes $c$ and $b$, respectively.

assumptions (R) and (F). The quantum switch, in particular, can maximally violate the VBC inequality (*53*)

$$P(a_2 = x_1, b = 0 | y = 0) + P(a_1 = x_2, b = 1 | y = 0) + P(b \oplus c = yz | x_1 = 0, x_2 = 0) \le \frac{7}{4} \quad (1)$$

where the notation $P(\cdot)$ denotes the probability that the condition inside the parenthesis is satisfied.

Let us first see why the existence of a definite causal order implies the inequality (1). The l.h.s. of Eq. 1 can be split into two contributions. The first contribution $P(a_2 = x_1, b = 0 \mid y = 0) + P(a_1 = x_2, b = 1 \mid y = 0)$ is the probability that the two Alices win in a variant of the "Guess Your Neighbor's Input" game (*62*), to be played whenever Bob's setting is $y = 0$: In this variant, Alice 2 has to guess Alice 1's setting in coincidence with Bob getting outcome $b = 0$, and Alice 1 has to guess Alice 2's setting in coincidence with Bob getting outcome $b = 1$. This term can at most achieve the value of 1. The term $P(b \oplus c = yz \mid x_1 = 0, x_2 = 0)$, instead, is the probability that Bob and Charlie win in the Clauser-Horne-Shimony-Holt (CHSH) game (63), to be played whenever Alice 1 and Alice 2 have settings $x_1 = 0$ and $x_2 = 0$, respectively. For experiments subject to the laws of quantum theory, this term can at most achieve the value $\left(2+\sqrt{2}\right)/4$, corresponding to the Tsirelson bound (64).

Now, under the assumption of definite causal order, the two contributions to the l.h.s. of Eq. 1 cannot jointly attain their maximum values. To maximize the first contribution, Bob's outcome must be perfectly correlated with the causal order of the experiments performed in the laboratories of the two Alices, so that, if $b = 0$, a signal can be sent from Alice 1 to Alice 2 informing her of Alice 1's setting and vice versa for $b = 1$. For this to be possible, $b$ must be perfectly correlated with the hidden variable $\lambda$. But if $b$ is perfectly correlated with $\lambda$, then it can be predicted from the value of $\lambda$, and this predictability prevents any violation of the CHSH inequality, meaning that the second contribution to the VBC inequality is upper bounded by the local realistic value 3/4 (53). In summary, maximization of the first contribution implies that the second contribution is upper bounded by 3/4, and therefore their sum is upper bounded by 7/4. More generally, VBC showed that the upper bound of 7/4 holds for every probability distribution satisfying the three assumptions D, R, and F.

In stark contrast, VBC showed that the quantum switch enables a maximal violation of the inequality (1), using the protocol illustrated in Fig. 1B. In this protocol, the two Alices measure a target qubit T, initially in the state $|0\rangle$, in the computational basis $\{|0\rangle, |1\rangle\}$, obtaining outcomes $a_1$ and $a_2$, respectively. After obtaining outcome $a_1$ ($a_2$), Alice 1 (Alice 2) resets system T to the state $|x_1\rangle$ ($|x_2\rangle$). Overall, the operations of Alice $i$ (with $i \in \{1,2\}$) are described by quantum instruments with Kraus operators $M_{a_i|x_i} := |x_i\rangle\langle a_i|$.

The order in which Alice 1 and Alice 2 operate is coherently controlled by the state of a control qubit C, giving rise to the quantum switch (2, 4). The net result of the operations of the two Alices combined by the quantum switch is that the target and control system evolve jointly, undergoing a quantum instrument with Kraus operators

$$S_{a_1,a_2|x_1,x_2} := M_{a_2|x_2} M_{a_1|x_1} \otimes |0\rangle\langle 0| + M_{a_1|x_1} M_{a_2|x_2} \otimes |1\rangle\langle 1| \quad (2)$$

The control qubit C is initially entangled with another qubit B in Bob's laboratory, and the two qubits are prepared in the maximally entangled state $|\Phi^+\rangle = (|0\rangle\otimes|0\rangle + |1\rangle\otimes|1\rangle)/\sqrt{2}$. Last, Charlie and Bob perform the measurements that maximize the quantum violation of the CHSH inequality, with Bob measuring in the computational basis $\{|0\rangle, |1\rangle\}$ when his setting is $y = 0$.

It is easy to see that the above setup achieves a maximal violation of the VBC inequality. When $y = 0$, Bob's measurement collapses the system C to the state $|b\rangle$ corresponding to Bob's outcome. If $b = 0$, the control qubits ends up in the state $|0\rangle$, and Eq. 2 implies that Alice 2 performs her measurement on the output of Alice 1's operation, leading to the condition $a_2 = x_1$. Instead, if $b = 1$, the control qubit ends up in the state $|1\rangle$, and Alice 1 performs her measurement on the output of Alice 2's operation, leading to the condition $a_1 = x_2$. Hence, the two Alices always succeed in guessing each other settings in coincidence with the appropriate outcomes in Bob's lab, thereby achieving the condition $P\left(a_2 = x_1, b = 0 \mid y = 0\right) + P\left(a_1 = x_2, b = 1 \mid y = 0\right) = 1$.

At the same time, Bob and Charlie maximally violate the CHSH inequality whenever $x_1 = x_2 = 0$. With these settings, the outcomes of the two Alices are necessarily equal to 0: The Alice who measures first must find outcome 0 because the target system is in the state $|0\rangle$, while the Alice who measures second must find outcome 0 because the other Alice has reset the target system to the state $|0\rangle$. For outcomes $a_1 = a_2 = 0$ and settings $x_1 = x_2 = 0$, the Kraus operator $S_{a_1,a_2|x_1,x_2}$ in Eq. 2 is $|0\rangle\langle 0| \otimes I$ and leaves the control qubit unchanged. Hence, this setting guarantees that Bob and Charlie perform their measurements on the maximally entangled state $|\Phi^+\rangle$, which allows them to achieve a CHSH violation by the maximal amount $\left(2+\sqrt{2}\right)/4$. Together, the above protocol leads to a violation of the VBC inequality by the maximal amount $1 + \left(2+\sqrt{2}\right)/4$.

### Experimental implementation

In our setup, illustrated in Fig. 2, the control qubit is the polarization DoF of a photon belonging to a polarization-entangled photon pair, generated by pumping a type II cut periodically poled $KTiOPO_4$ (ppKTP) crystal in a Sagnac interferometer configuration (see Supplementary Materials for details). The target qubit, encoded in the time-bin DoF, is created by reshaping a continuous wave laser into periodic pulse trains, with a delay of 1 μs between the early and late time slots. The signal photon from the entangled pair is directed into the quantum switch, where operations are performed on its time-bin DoF at the local stations of Alice 1 and Alice 2. After traversing the switch, the photon's polarization is measured by Charlie. The idler photon is transmitted directly to Bob's station via an optical fiber, which we choose to be 3 km long to mimic the condition of spacelike separation during each run of the experiment.

In the quantum switch, the causal order of the two operations acting on the target qubit is determined by the state of the control qubit. In our case, when the polarization is in the horizontal polarization state $|H\rangle$, the photon is transmitted at a polarizing beam splitter (PBS1 in Fig. 2) and sequentially passes through Alice 1 and then Alice 2, where a measurement-and-repreparation process is applied to the photon's time-bin qubit. Conversely, when the polarization is in the vertical polarization state $|V\rangle$, the photon is reflected at PBS1, thereby passing through Alice 2 first and then through Alice 1. In our experiment, the control qubit of the quantum switch is maximally entangled with the qubit shared with Bob. When Bob projects his photon onto the computational basis, the control qubit collapses to a computational-basis state, and the target qubit undergoes operations in a definite causal order. In contrast, when Bob performs a diagonal-basis measurement, the control qubit collapses to a diagonal state, resulting in an indefinite causal order for the target qubit. Figure 2 illustrates the photon trajectories inside the quantum switch for control-qubit states $|H\rangle$ and $|V\rangle$, marked with time-stamped arrows.

In each run of the experiment, the target qubit is initialized in the early time-bin state $|e\rangle$, while the control qubit is entangled with Bob's system, and the two systems are prepared in the Bell state $(|HH\rangle + |VV\rangle)/\sqrt{2}$. The entangled pairs are generated by a 150-mW continuous wave pump laser, which produces approximately 2,000,000 pairs of entangled photons per second. After shaping the pump beam into a pulse train with a pulse width of 600 ns and a repetition rate of 50 kHz, the photon pair generation rate is reduced to 60,000 and further drops to 1400 after transmission through the quantum switch. The repetition rate of the pump laser matches the operating speed of the time-bin qubit and ensures that only one pulse enters the quantum switch per experimental cycle, preventing overlap between successive trials.

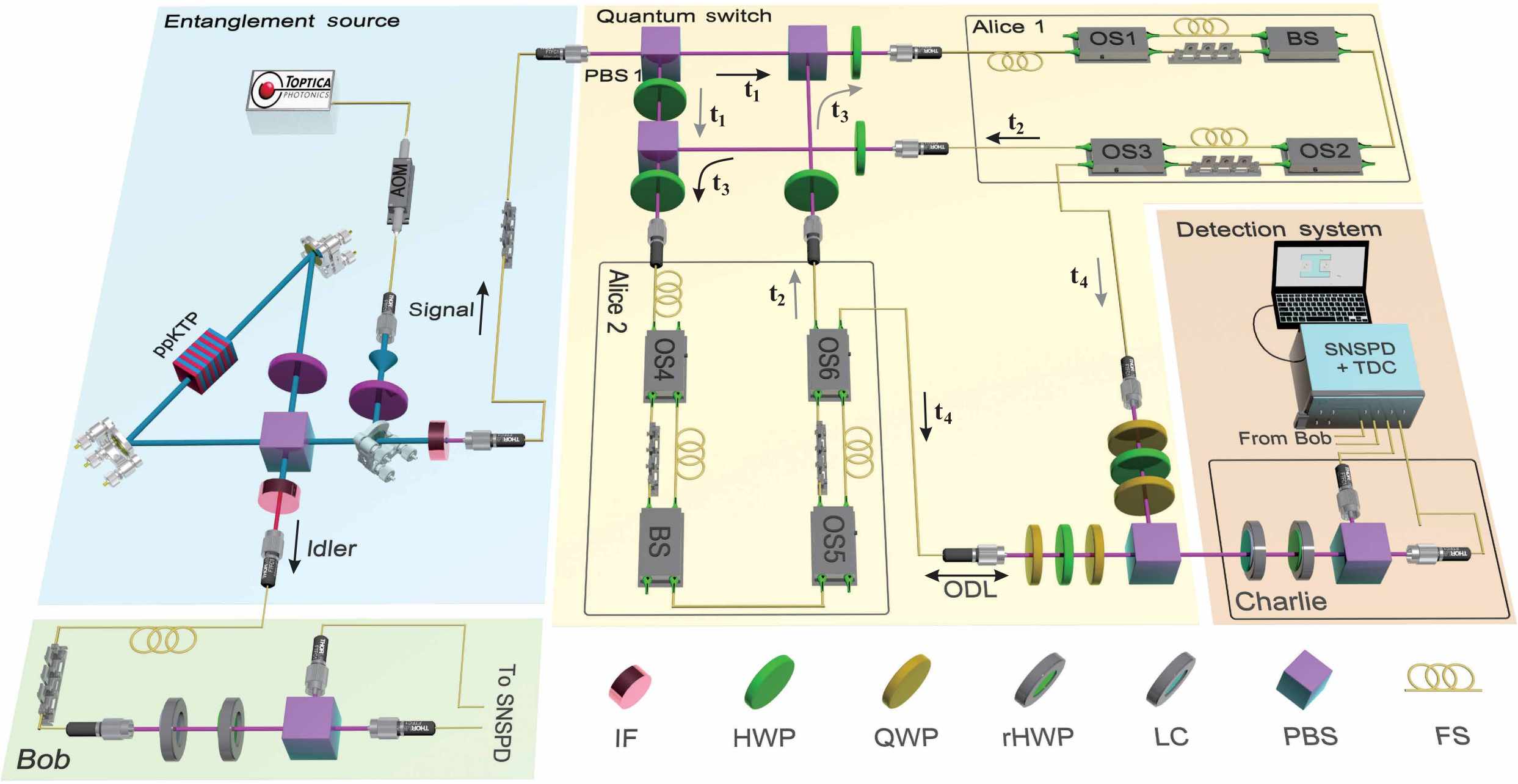

**Fig. 2. Optical layout of the experimental setup.** A 150-mW continuous wave laser at 775 nm are reshaped into periodic pulses with an acousto-optic modulator (AOM), which pumps a type II cut ppKTP crystal in a Sagnac configuration, effectively generating entangled photon pairs at 1550 nm (light blue). The signal photon is guided into the quantum switch (yellow) and then measured by Charlie (rectangle, light red), while the idler photon is distributed to Bob directly (green). In the quantum switch, the photons' polarization and time-bin serve as the control qubit and target qubit, respectively. The flow of photons is indicated by arrows with time labels ($t_1$ to $t_4$), where black and gray arrows represent the control qubit being in state $|H\rangle$ and state $|V\rangle$, respectively. In each Alice's local station (rectangle, yellow), a measure-and-reprepare operation on the time-bin in implemented by using electro-optical switches (OSs), beam splitter (BS), and polarization controllers. An optical delay line (ODL) and a liquid crystal (LC) variable retarder are used to set the path length and the relative phases of the interferometer. The electrical control system, including master oscillator, RNGs, and field-programmable gate arrays (FPGAs), is shown in the Supplementary Materials. IF, interference filter; HWP, half-wave plate; rHWP, HWP with motorized rotation stages; QWP, quarter-wave plate; PBS, polarizing beam splitter; FS, fiber spool; SNSPD, superconducting nanowire single-photon detector.

To test the VBC inequality, Alice $i \in \{1,2\}$ has to project the target qubit in the computational basis, thereby generating the outcome $a_i$. She then reprepares the target qubit in the basis state $|x_i\rangle$ according to her input $x_i$. Afterward, Charlie measures the control qubit using the observable $Z + X$ when $z = 0$, or $Z - X$ when $z = 1$, and records the outcome $c$. Simultaneously, Bob measures the idler photon in the $Z$ basis for $y = 0$, or in the $X$ for $y = 1$, and records the outcome $b$. The resulting statistics are used to estimate the joint probability distribution $P\left(a_1, a_2, b, c \,|\, x_1, x_2, y, z\right)$.

The measurements performed by Alices 1 and 2 are implemented using an asymmetric Mach-Zehnder interferometer (AMZI) comprising an electro-optic (EO) switch, a beam splitter, optical fibers, and a polarization controller. The path length difference of the AMZIs is 200 m, matching the temporal separation defined for our time-bin qubit. For state repreparation, the beam splitter is replaced with an EO switch (OS3 or OS6 in Fig. 2), allowing photons to be dynamically routed either between Alice 1 and Alice 2 or out of the quantum switch. To guarantee that the repreparation settings of Alice 1 and Alice 2 are chosen independently and randomly in each experimental round, we use two RNGs to drive switches OS2 and OS5. Polarization controllers, along with assemblies of quarter-, half-, and quarter-wave plates, are used to preserve polarization across all fiber loops.

An important point is that the state repreparation is independent of the measurement outcomes $a_i$: Alice 1 and Alice 2 only need to reset the target qubit to the state corresponding to the value of their settings $x_i$. Thanks to this fact, the readout of the outcomes can be delayed until after the control qubit has been measured, thereby avoiding that the order in which the measurement outcomes are produced reveals the order in which the operations of Alice 1 and Alice 2 took place, reducing the superposition of orders to a classical mixture. In our experiment, this delayed-readout strategy is implemented by introducing an ancillary time-bin qubit together with several electrical trigger signals that encode the measurement outcomes and repreparation settings $\{a_i, x_i\}$ in the time domain. These signals are subsequently decoded using a TDC. Notably, our configuration avoids the need for duplicated optical components required by a previous technique based on the polarization DoF (*43*). See Materials and Methods for more details.

For Bob's and Charlie's measurements on the polarization DoF, we use a half-wave plate (HWP) and a PBS to implement the observables required by the VBC inequality. The HWPs are mounted on motorized rotation stages, allowing the measurement bases to be set remotely. While our current setup uses motorized control, faster polarization measurements can be achieved using EO modulator–based techniques (*55*, *56*). After the PBSs, photons are coupled back into fibers and detected by superconducting nanowire single-photon detectors. The detection signals are processed by the TDC, which records the photon arrival times.

A key requirement for a VBC test is the implementation of fast operations performed in a coordinated way on the DoFs associated to Alices', Bob's, and Charlie's systems. This requirement makes VBC tests more challenging compared to standard Bell tests, where fast control of only a single DoF—typically polarization—is sufficient. For loophole-free tests of Bell nonlocality, repetition rates of up to 1 MHz have been demonstrated (*55*, *56*). In our experiment, we implement operations on the time-bin DoF at a repetition rate of 0.1 MHz, which can be further improved by suitable enhancements in our setup, as discussed in the Supplementary Materials. Precise synchronization is essential throughout the experiment; otherwise, photons may be misrouted by the EO switches. To ensure proper timing, we lock the driving signals for the acousto-optic modulator, EO switches, RNGs, and the TDC trigger to a common 10-MHz master oscillator. These driving signals are generated using field-programmable gate arrays. In summary, the techniques developed in our experiment guarantee that the operations performed by the two Alices are sufficiently fast to achieve spacelike separation with Bob at the distance simulated by the fiber spools used in our experiment. Incorporating existing state-of-the-art fast polarization measurements at Bob's and Charlie's measurement stations, together with our current configuration, would then attain all the hardware requirements for guaranteeing spacelike separation in the VBC test.

## Experimental results

The theoretical values for each term in Eq. 1 are $\frac{1}{2}$, $\frac{1}{2}$, and $\frac{1}{2}+\frac{\sqrt{2}}{4}$, respectively. The values measured in our experiment are $0.490 \pm 0.004$, $0.492 \pm 0.004$, and $0.825 \pm 0.009$, close to the theoretical predictions. These values lead to an experimental value of the left side of the VBC inequality equal to $1.807 \pm 0.010$, shown in Fig. 3. The measurement settings are implemented independently for all agents and the space-time relations among them are set according to Fig. 1, justifying the assumptions R and F of the VBC inequality. Hence, any explanation in terms of fixed causal structure of these agents is excluded by 5.7 SDs.

The value of the VBC inequality observed in our experiment is approximately 0.05 lower than the theoretical value. We now provide a brief error analysis. Two main factors contribute to this discrepancy: First, bit-flip noise arising from cross-talk in the EO switches and polarization disturbances in the long optical fibers. The optical switch used in our experiment has an isolation of 20 to 23 dB, which corresponds to a cross-talk range between 1 and 0.5%. Second, phase noise occurs in the interferometer due to temperature fluctuations. We implemented temperature control to maintain an interference visibility of 0.98 over a 15-min period, which minimized the phase decoherence caused by temperature variations, although some residual effects still persist (see Materials and Methods). Specifically, for the first two terms of the inequality, when Bob measures in the computational basis, the target qubit of the quantum switch undergoes the operations of the two Alices in a definite causal order. The primary source of error here is bit-flip noise. For the third term, when Bob measures in the diagonal basis, the quantum switch forms a Mach-Zehnder (MZ) interferometer. In this case, both bit-flip and phase-flip noise contribute to the deviation of the experimental value from the theoretical value.

We further test the no-signaling conditions implied by the light-cone structure in the VBC scenario, checking that a party's outcome cannot be affected by the setting of parties in their causal future or spacelike separated parties. The relevant constraints are detailed in the Supplementary Materials. For example, when Bob projects his qubit onto the state $|H\rangle$, the target qubit of the quantum switch undergoes operations first by Alice 1 and then by Alice 2. In this case, the measurement outcome of Alice 1 must be independent of Alice 2's repreparation choice $x_2$. To test this assumption, we calculate $\delta P(a_1 | x_2, y=0, b=0) := P(a_1=0 | x_2=0, y=0, b=0) - P(a_1=0 | x_2=1, y=0, b=0)$ as a measure of the effect of $x_2$ on $a_1$ when Bob measures in the computational basis and finds outcome 0, with the corresponding result shown as the first data point in Fig. 4. The figure summarizes the outcomes of all no-signaling checks, showing that the number of probability differences falling within 1 SD, between 1 and 2 SDs, and between 2 and 3 SDs are 5, 4, and 3, respectively, regardless of the measurements performed on the causal future or by spacelike separated agents. For a normally distributed event, the probability of all 12 independent replication samplings yielding results within 3 SDs is ~96.8%, while the probability of at least one sampling exhibiting more than 4 SDs is less than 0.1%. This confirms that our experiment satisfies the no-signaling constraints associated to the VBC scenario with high statistical significance. Last, it is worth mentioning that our experimental setup also permits the testing of additional DRF inequalities introduced in (*53*) (see the Supplementary Materials for further discussion).

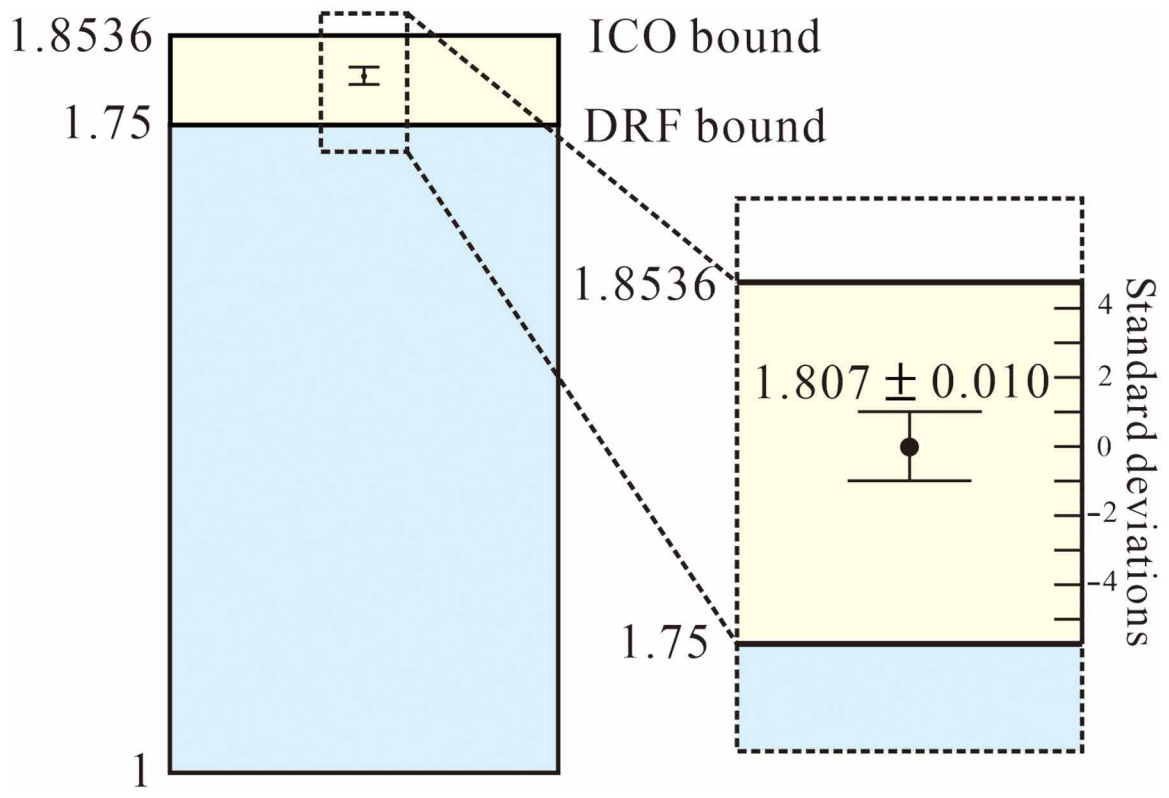

**Fig. 3. Experimental data for the VBC inequality test.** The observed value for Eq. 1 is 1807 ± 0.010, while the DRF bound and the ICO bound are 1.75 and 1.8536, respectively.

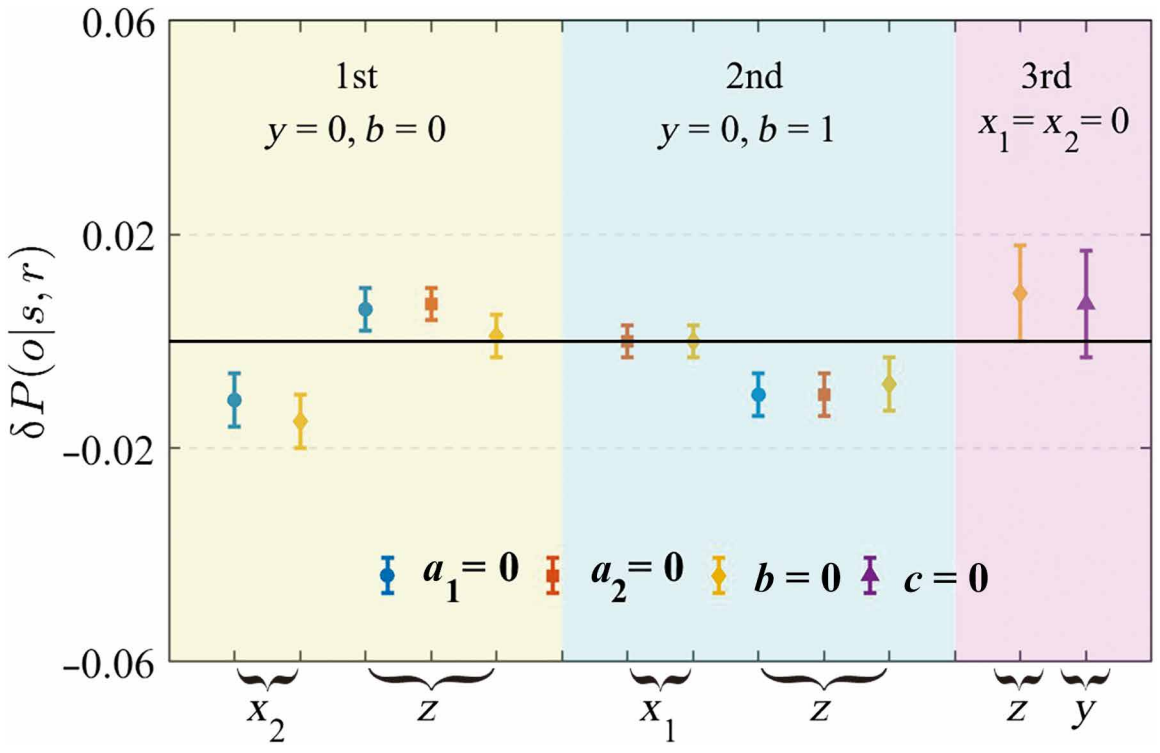

**Fig. 4. Results of no-signaling test.** Here, we test the no-signaling constraints implied by the causal structure in Fig. 1A. The potential signaling is quantified by probability differences $\delta P(o=0 | s, r) := P(o=0 | s=0, r) - P(o=0 | s=1, r)$, which measure the influence of a setting $s \in \{x_1, x_2, y, z\}$ (specified by the curly brackets along the horizontal axis) on an outcome $o \in \{a_1, a_2, b, c\}$ (corresponding to the blue, orange, yellow, and purple data points in the figure), for fixed values of some of the remaining settings/outcomes $r$, specified by the shaded areas in yellow ($y=0, b=0$), blue ($y=0, b=1$), and pink ($x_1=x_2=0$). For example, the first data point corresponds to $\delta P(a_1=0 | x_2, y=0, b=0)$, obtained by marginalizing $p(a_1, a_2, b, c | x_1, x_2, y, z)$ over the variables $a_2$ and $c$, by averaging over the possible values of $z$ and $x_1$, and by conditioning on $b=0$, corresponding to the causal order in which Alice 1 precedes Alice 2. In this case, $\delta P(a_1=0 | x_2, y=0, b=0)$ quantifies the influence of $x_2$ on $a_1$, on average over the remaining settings/outcomes. In general, for all the terms shown in the figure the values of δP lie within 3 SDs of zero, confirming that the no-signaling conditions are well satisfied.

## DISCUSSION

Here, we experimentally demonstrated a violation of the VBC inequality, exhibiting correlations that are not compatible with the existence of a definite causal order between the operations performed by two parties, Alice 1 and Alice 2, under the standard assumptions that relativistic causality holds and free interventions are possible. Overall, the experimental advancements achieved to demonstrate the VBC inequality, together with those reported in recent independent studies (*65*, *66*), provide a valuable tool for foundational explorations on quantum causality, as well as for applications to quantum information processing.

It is important to stress, however, that the violation of the VBC inequality reported in our experiment is not loophole free. First, similar to standard Bell tests, our demonstration is subject to the locality loophole, the fair sampling loophole, and the free-choice loophole. These loopholes can be closed with hardware modifications to our experiment, incorporating the techniques established in recent loophole-free tests of Bell inequalities (*55*–*57*). An additional loophole, however, arises from the delayed measurement strategy used in our experiment: The operations of Alice 1 and Alice 2 take place over extended periods of time, during which a signal could in principle be sent from the laboratory of Alice 1 to the laboratory of Alice 2 and vice versa. This situation would break one of the assumptions used in the derivation of the VBC inequality (assumption D), which assumed a strict dichotomy: Either Alice 1 causally precedes Alice 2 or vice versa. By breaking this assumption, one could in principle reproduce the violation of the VBC inequality by a model where the two Alices communicate their settings to each other during the execution of their operations: In this way, the two Alices could always guess each other's settings without errors, and, independently, Bob and Charlie could perform a Bell experiment that maximally violates the CHSH inequality, thereby achieving the maximal quantum violation of the VBC inequality.

In our setup, bidirectional signaling between Alice 1 and Alice 2 is effectively ruled out by the observation that the causal influences between the laboratories of the two Alices are mediated by a single photon, which only travels from the laboratory of one Alice to the laboratory of the other Alice in each branch of the interferometer. This argument, however, is not device independent, because it relies on knowledge of the setup. It is worth observing that this loophole is not specific to our experiment: It is common to all photonic implementations of the quantum switch and, more generally, to all implementations in which the operations of the two Alices admit a fine-grained description unfolding in classical spacetime without causal loops (*9*). Nevertheless, this loophole could be possibly lifted in future experiments probing the quantum nature of spacetime (*67*–*69*), provided that one can find spacetime scenarios in which bidirectional signaling can be ruled out at a fundamental level. Whether these implementations are possible and whether a device independent, loophole-free certification of indefinite causal order is in principle possible in nature is a profound question with fundamental implications for quantum mechanics and spacetime physics and, more broadly, for the very notion of causality.

## MATERIALS AND METHODS

### Readout of time-bin qubit measurement results in the quantum switch

In our experiment, extracting the measurement outcomes from within the quantum switch poses a substantial technical challenge. Since the system qubit is measured by Alice 1 and Alice 2 inside the switch, most standard readout methods would reveal which operation occurred first, thereby destroying the quantum superposition of causal orders. To circumvent this problem, we adopt the delayed-readout approach introduced in (*43*), in which the extraction of measurement results is postponed until after the control qubit has been measured.

In general, the delayed-readout approach of (*43*) requires $d$ copies of the measurement apparatus, where $d$ is the total number of outcomes associated with the interactions inside the quantum switch. This requirement arises because measurements on a polarization qubit generate additional spatial modes, which cannot be coherently erased until the outcomes are read out. In our implementation, however, we overcome this limitation by introducing an ancillary time-bin qubit and encoding the measurement outcomes in the temporal domain rather than in additional spatial modes. Specifically, we coherently control the propagation of the time-bin qubit using EO switches embedded in AMZIs. By introducing an additional 200-m fiber delay in three of the four AMZIs, the measurement outcomes are mapped onto distinct arrival times, allowing them to be deterministically distinguished by the detectors and the TDC. Furthermore, to preserve the information about the repreparation choices of Alice 1 and Alice 2, we duplicate the driving signals of OS2 and OS5 and use them as trigger signals for the TDC. In this way, the combinations of measurement outcomes and repreparation settings $\{a_i, x_i\}$ can be labeled directly in the temporal domain without introducing extra spatial paths. More details can be found in the Supplementary Materials.

### Phase stabilization via temperature control

A major challenge in our setup is maintaining phase stability in the MZ interferometer embedded in the quantum switch, which is highly sensitive to environmental perturbations, especially temperature fluctuations. These fluctuations, primarily caused by variations in the refractive index of the fiber and air, lead to fluctuations in the optical path lengths of the two interferometer arms. Given the fast-switching loops involved in Alices' operations on the time-bin DoF, implementing an active phase-locking system—like those used in previous experiments (*21*, *32*)—is impractical.

To mitigate phase drift, we use two strategies. First, we maximize the overlap of the optical paths between the two causal orders, ensuring that the length of unshared fiber segments is less than 20 cm and the free-space length is less than 10 cm. This free space also serves as an optical delay line to compensate for differences in fiber length. Second, we designed and implemented a temperature control system to keep fluctuations below 0.05°C, stabilizing the entire tabletop setup. We monitored the phase stability of the MZ interferometer for a 15-min period. With an input of 1.5 mW and the interferometer set to destructive interference, the averaged output power, shown in Fig. 5, is ~15.2 μW, corresponding to an overall visibility of 0.980.

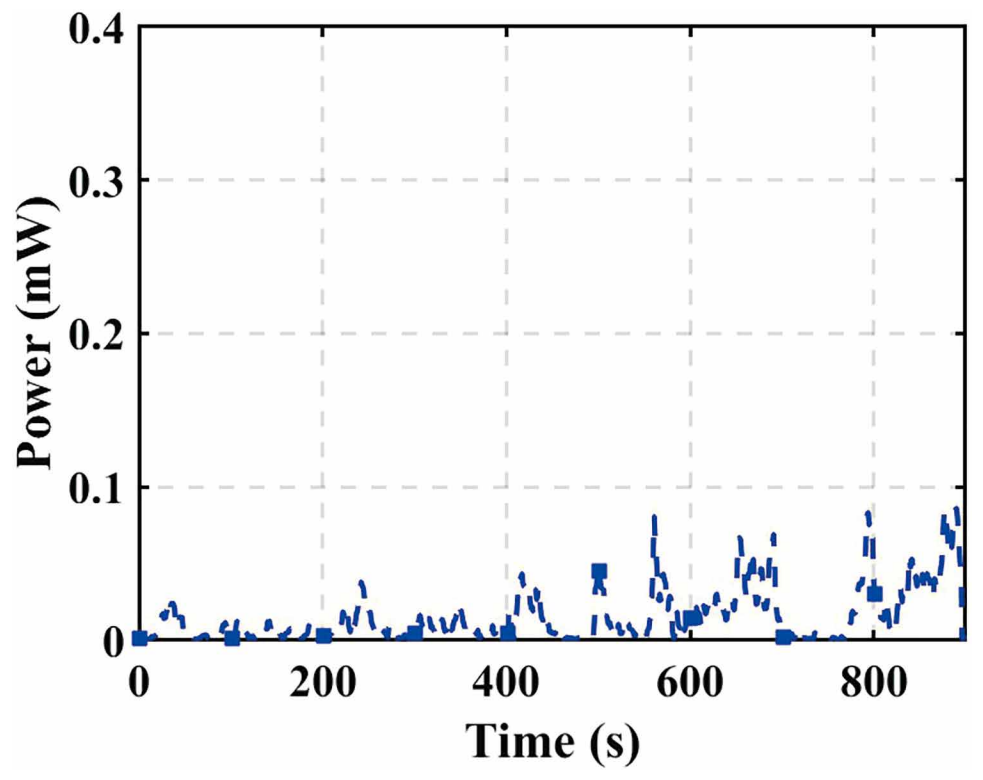

**Fig. 5. Observed output power of the Mach-Zehnder interferometer in our setup over 15 min.** The input power is 1.5 mW, and the output power is recorded once per second. The averaged power at destructive interference is about 15.2 μW during this period.

## Supplementary Materials

**This PDF file includes:**

Supplementary Text

Figs. S1 to S3

Table S1

**Acknowledgments:** We thank G. Chen for stimulating discussions and valuable comments. **Funding:** The USTC group was supported by the NSFC (nos. U25D8007, 62322513, 12374338, 12574402, and 12350006), the Quantum Science and Technology-National Science and Technology Major Project (no. 2021ZD0301200), the Fundamental Research Funds for the Central Universities (WK2030250138 and WK2030250119), Anhui Provincial Natural Science Foundation (no. 2408085JX002), Anhui Province Science and Technology Innovation Project (no. 202423r06050004), Guizhou Provincial Major Scientific and Technological Program XKBF (2025)010, 009, and Xiaomi Young Scholar. G.C. was supported by the Chinese Ministry of Science and Technology through grant 2023ZD0300600 and by the Hong Kong Research Grant Council through the Senior Research Fellowship Scheme SRFS2021-7S02 and the Research Impact Fund Scheme R7035-21F. **Author contributions:** B.-H.L., G.C., C.-F.L., and G.-C.G. conceived the project and supervised the research. Y.G., H.T., and B.-H.L. designed the experimental proposal. H.T. and Y.G. performed the experiment with the help of B.-X.W., M.-Y.L., J.-W.F., X.-M.H., and Y.-F.H. H.T. and Y.G. analyzed the experimental data. G.C., Y.G., and B.-H.L. wrote the manuscript with the help of other authors. **Competing interests:** The authors declare that they have no competing interests. **Data, code, and materials availability:** All data and code needed to evaluate and reproduce the results in the paper are present in the paper and/or the Supplementary Materials. This study did not generate new materials.

Submitted 27 November 2025
Accepted 29 April 2026
Published 10 June 2026
10.1126/sciadv.aee2912

# Experimental violation of a Bell-like inequality for causal order

Yu Guo, Hao Tang, Bo-Xuan Wang, Min-Yu Lv, Jia-Wen Fan, Xiao-Min Hu, Yun-Feng Huang, Chuan-Feng Li, Guang-Can Guo, Giulio Chiribella, and Bi-Heng Liu

# Supplementary Materials for

## Experimental violation of a Bell-like inequality for causal order

Yu Guo *et al.*

Corresponding author: Giulio Chiribella, giulio@cs.hku.hk; Bi-Heng Liu, bhliu@ustc.edu.cn

**This PDF file includes:**

Supplementary Text
Figs. S1 to S3
Table S1

# 1 Core experimental innovations

For completeness, we summarize the main experimental innovations of our work in this section.

1. *Quantum switch with polarization control and time-bin target.* Our experiment realizes, for the first time, a photonic quantum switch in which a polarization-encoded control qubit coherently controls operations acting on a time-bin target qubit. Previous optical demonstrations of the quantum switch have typically employed other combinations of degrees of freedom, such as path-controlled polarization, polarization-controlled transverse-electric modes, or path-controlled time-bin qubits. In contrast, our implementation combines polarization control with a time-bin target. This encoding offers a practical advantage: both polarization and path operations can be manipulated rapidly using electro-optic elements, making the architecture particularly well suited to scenarios—such as the present experiment—in which fast operations are required on both the control and target systems. This implementation enriches the experimental toolbox for realizing indefinite causal order in photonic platforms.

2. *Implementation of a measurement-and-repreparation process on a time-bin qubit with delayed readout of the measurement outcomes.* Our experiment implements a time-bin–based strategy to realize the measurement-and-repreparation protocol inside a quantum switch without generating additional spatial modes. By embedding electro-optic (EO) switches in asymmetric Mach–Zehnder interferometers (AMZIs), we coherently manipulate an ancillary time-bin qubit and encode the measurement outcomes in the temporal domain. The outcomes are mapped onto distinct photon arrival times through additional fiber delays, while the repreparation settings are recorded via trigger signals synchronized with the EO switch driving signals. This approach enables deterministic readout of the mesurement outcome and repreparation choice combinations directly in the time domain, avoiding the need for multiple copies of the measurement apparatus required in polarization-based implementations.

3. *Phase stabilization of a kilometer-scale Mach–Zehnder interferometer via temperature control.* Our experiment further develops a temperature-controlled stabilization scheme to ensure stable interference in the quantum switch Mach–Zehnder interferometer under fast time-bin operations. Instead of active phase locking—which is difficult to implement with rapidly switching optical paths—we minimize the length of the unshared optical paths between the two causal orders and stabilize the entire experimental platform using a custom temperature control system. This design suppresses phase drift induced by environmental fluctuations and enables a long-term interference visibility of about 0.98, which is crucial for observing the violation of the DRF inequality.

## 2 Details on the entangled photon source

The experiment reported in this work was conducted with a photonic polarization entangled source based on a spontaneous parametric down-conversion process on a type-II cut ppKTP crystal in a Sagnac configuration. A 150 mW diode laser centered at 775 nm was reshaped into a pulse array with a acoustic optical modulator (AOM). The pulse width and repetition rate were 600 ns and 50 kHz, respectively. Setting the polarization of the pump laser to be diagonal, we generated pairs of entangled photons centered at 1550 nm in a polarization state $(|H\rangle|V\rangle + |V\rangle|H\rangle)/\sqrt{2}$, which were then separated by a polarizing beam splitter (PBS). The pump laser was blocked with long-pass and narrow-band pass filters. Then, the entangled photon pairs were coupled into single-mode fibers and transformed into $(|H\rangle|H\rangle + |V\rangle|V\rangle)/\sqrt{2}$ with a fiber polarization controller. The idler photon was distributed to Bob while the signal photon was guided into the quantum switch and finally measured by Charlie. When measuring the photon pairs in the diagonal basis, we observed a visibility of more than 0.99, indicating a good quality of our entangled source.

The duty cycle of our pump pulses was 3% and the threshold power of the AOM we used is 200 mW, which limits the pump power of our entanglement source to less than 6 mW. When setting the coincidence window to 0.5 ns, the observed coincidence rate of the photon source was about 60,000 pairs per second, the counting rate of each detector was about 180,000, and thus the coincidence efficiency was 0.33. The coincidence rate was attenuated to 1400 pairs per second after the signal photon passed through the whole apparatus. The drop in collection rate with and without the quantum switch indicates that the switch introduces a loss of more than 16 dB. This loss is primarily due to three factors: the insertion loss of the optical switch (approximately 1.5 dB each), the fiber beam splitters (3 dB each), and the losses associated with fiber-to-free-space coupling (0.5 dB each).

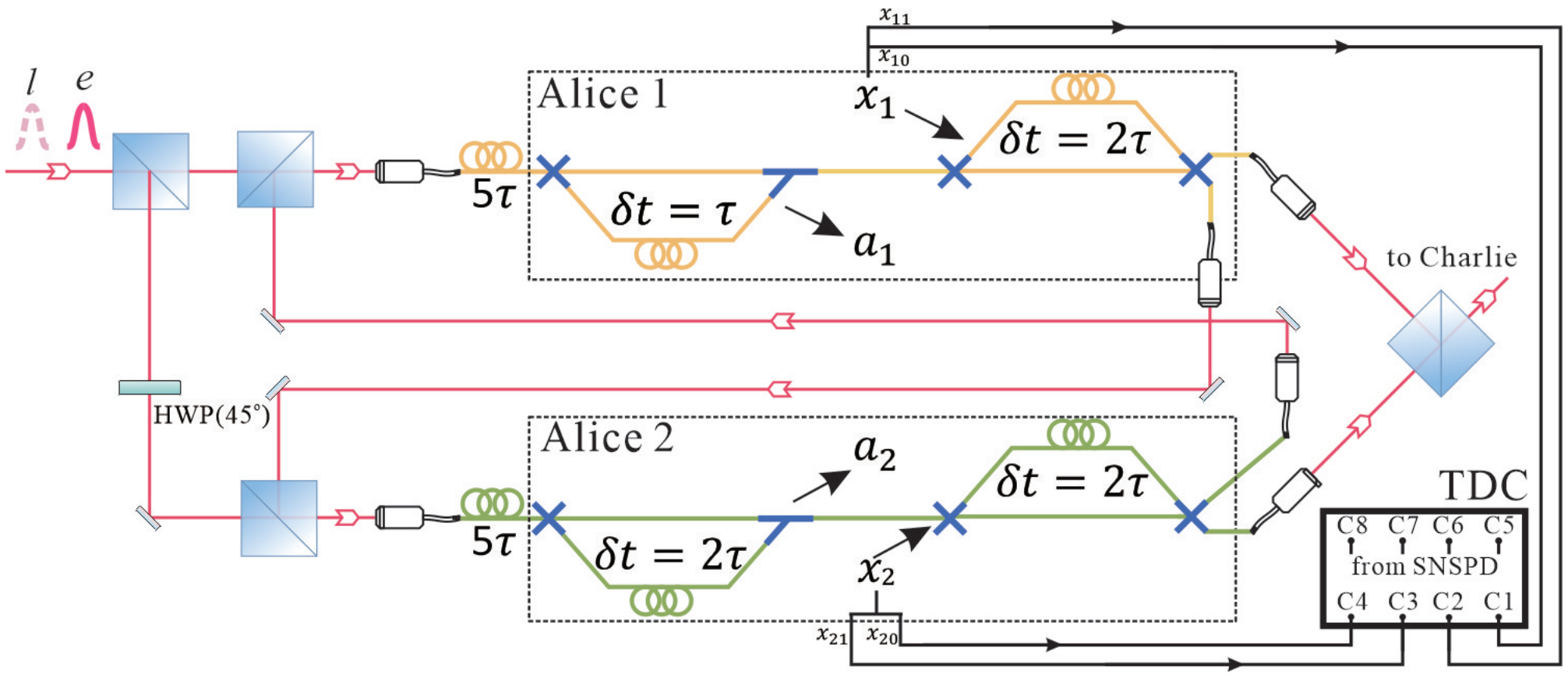

**Figure S1**: **The reading out of $a_i$ and $x_i$.** For each Alice, the first asymmetric Mach–Zehnder interferometer (AMZI) is used to perform a measurement on the time-bin target qubit, while the second is used to reset the target state. The time-bin qubit is defined by a relative time delay of $\tau = 1\ \mu$s. To enable the retrieval of the measurement outcomes $a_i$ without disturbing the control qubit, an additional time delay $\tau$ is introduced in all AMZIs except for the measurement AMZI of Alice 1. To retrieval the repreparation choices $x_i$, four trigger signals (amplitude 3.3 V, width 10 ns), synchronized to the resetting signal $x_i$ (the driving signals of OS2 and OS5 in Fig. 2 of the main text), is guided to four channels (C1-C4) of the TDC. The other four TDC channels (C5-C8) are used to record the detected photon events from the SNSPD. TDC, time-to-digital converter; SNSPD, superconducting nanowire single-photon detector.

# 3 Details on the time-bin encoding and the reading out of $a_i$ and $x_j$

Time-bin encoding constitutes a robust photonic qubit encoding scheme, wherein the logical basis states are defined by the photon arrival time in two well-separated temporal modes, commonly denoted as the early and late time bins. Such qubits can be generated and coherently manipulated using AMZIs, where the path-length difference between the two arms imposes a fixed temporal delay between the temporal modes. Here the path-length difference is required to be significantly larger than the single photon coherence length to ensure that the early and late bins are well separated in the time domain with negligible temporal overlap. By routing photons through the short or long arm of an AMZI, one can map quantum information onto the temporal degree of freedom while preserving phase coherence. To measure the time-bin qubit, a second AMZI is typically used to project the temporal superposition onto the target logical basis, followed by post selecting the middle time bin detecting events. Importantly, the measurement AMZI is configured such that the early bin from the preparation stage propagates along the long arm and the late bin along the short arm, which compensates the temporal delay and enables coherent interference.

In our implementation, the time-bin qubit is defined by a relative time delay $\tau = 1$ $\mu$s, corresponding to a path length difference of 200 meters between the two fiber arms of the AMZI. The operations on the time-bin qubit are performed at a repetition rate of 0.1 MHz. This rate is primarily limited by the performance of the EO switches. The EO switches used in our experiment have a rising time of less than 100 ns at a switching rate of no more than 0.3 MHz. While its isolation reaches about 30 dB at low switching rates (a few Hz), this isolation degrades as the switching rate increases. At our operating rate of 0.1 MHz, the isolation was measured to be 20-23 dB. Another factor that restricts the operation rate is the overall optical loss of the system. To maintain a sufficiently high counting rate of the photon source, we employed a moderate pump pulse duty cycle. Specifically, we set the pulse width to 600 ns in our experiment. The use of high-performance commercial optical switches (e.g. BATi Nanona fiber switches, rising time $<$ 60 ns, repetition rate up to 1 MHz) could significantly enhance the operation speed of time-bin qubit manipulations, potentially by an order of magnitude.

In our experiment, time-bin encoding plays a central role in implementing the measurement-and-repreparation protocol within the quantum switch, as it allows measurement outcomes $a_i$ and control settings $x_i$ to be recorded in the temporal domain and retrieved at a later stage without disturbing the coherence of the control qubit. To this end, we introduce an ancillary time-bin qubit that defers the readout of measurement results until after the control qubit has been measured.

Specifically, each Alice employs two AMZIs: the first projects the target qubit onto the computational basis to yield the outcome $a_i$, and the second resets the target qubit in the state $|x_i\rangle$ according to the input $x_i$. In the measurement AMZI, photons propagating through the long arm correspond to $a_i = 0$, while those passing through the short arm correspond to $a_i = 1$. Conversely, in the repreparation AMZI, routing a photon into the long arm prepares the target qubit in the state $|1\rangle$, whereas the short arm corresponds to preparation in $|0\rangle$. As illustrated in Fig. S1, we introduce an ancillary time-bin qubit by imposing an additional time delay $\tau$ in three of the four AMZIs. The ancillary qubit expands the dimension of the Hilbert space, enabling the measurement outcomes of Alice 1 and Alice 2 to be mapped into sequential distinct temporal modes; the relative delay of the detected signals thus reveals the measurement outcome $a_i$.

In parallel, we use four trigger pulse signals $x_{ij}$ (($i \in \{1,2\}, j \in \{0,1\}$)) to record the repreparation setting choices of Alice 1 and Alice 2. Here, $x_{ij}$ denotes the setting choice $j$ of Alice $i$ and is used to trigger four channels of the time-to-digital converter (TDC, C1-C4 of Fig. S1). The amplitude and width of these $x_{ij}$ signals are set to 3.3 V and 10 ns to satisfy the operating conditions of the TDC. To maintain precise temporal synchronization, the $x_{ij}$ signals are phase-locked to the driving signals of the EO switches. Specifically, $x_{10}$ and $x_{11}$ are synchronized to the low and high levels of the driving signal for OS2, respectively, while $x_{20}$ and $x_{21}$ are synchronized to the low and high levels of the driving signal for OS5. Here, OS2 and OS5 denote the first EO switches in the repreparation AMZIs of Alice 1 and Alice 2, respectively (Fig. 2 of the main text). Under these configurations, all signals $a_i$s and $x_i$s can be deterministically distinguished via the combination of the trigger channel and the relative time delay between the two-photon detection events, as summarized in Tab. S1.

Note that the length of the fibers in the AMZI can be measured with a fluctuation of less than 2 cm. This precision is achieved by comparing the photon arrival times with and without the fiber inserted. Given that the pulse width in our experiment is 600 ns, a 2 cm variation in fiber length is negligible.

# 4 Timing coordination and spacetime configuration in the experiment

As discussed in the main text, the causal relations among the operations of the four agents are crucial for our experiment. According to Fig. 1(A) of the main text, Alice 1 and Alice 2 lie in the causal past of Charlie, while Bob is spacelike separated from the other three parties. To satisfy these causal relations, both the spatial separations between the agents and the durations of their individual operations must be carefully considered. Except for a 10 cm free-space optical delay line, the four agents are connected via fiber spools. The idler photon is sent to Bob through a 3 kilometer fiber, corresponding to a delay of 15 $\mu$s. Inside the quantum switch, a 1 kilometer fiber is placed before each Alice's laboratory. In addition, the AMZIs have short arms consisting of 1 meter fibers and long arms consisting of 201 meter and 401 meter fibers, respectively. The fiber configuration is shown in Fig. S1, where the unlabeled links are typically 1 meter fibers. In the following discussion, we neglect the delay introduced by these 1 meter fibers, as their propagation delay (about 5 ns) is much smaller than the temporal separation defining our time-bin qubit. Under these conditions, the maximal total length of the path traveled by the signal photon inside the quantum switch is approximately 3.4 kilometers, corresponding to a delay of about 17 $\mu$s.

**Table S1**: The time delay and trigger channel settings to read out $a_1x_1a_2x_2$.

| $a_1x_1a_2x_2$ | relative time delay | trigger detector | $a_1x_1a_2x_2$ | relative time delay | trigger detector |
|---|---|---|---|---|---|
| 0000 | $3\tau$ | D1D3 | 1000 | $2\tau$ | D1D3 |
| 0001 | $5\tau$ | D1D4 | 1001 | $4\tau$ | D1D4 |
| 0010 | $\tau$ | D1D3 | 1010 | 0 | D1D3 |
| 0011 | $3\tau$ | D1D4 | 1011 | $2\tau$ | D1D4 |
| 0100 | $5\tau$ | D2D3 | 1100 | $4\tau$ | D2D3 |
| 0101 | $7\tau$ | D2D4 | 1101 | $6\tau$ | D2D4 |
| 0110 | $3\tau$ | D2D3 | 1110 | $2\tau$ | D2D3 |
| 0111 | $5\tau$ | D2D4 | 1111 | $4\tau$ | D2D4 |

To satisfy these causal relations, a high level of coordination between the optical and electronic subsystems is required. This coordination, as illustrated in Fig. S2, relies on the precise synchronization of all optoelectronic components, including the AOM, random number generators (RNGs), and EO switches. To achieve this, the operating clocks of these devices are synchronized throughout the system. The AOM is driven by a digital signal generator (DSG), while all EO switches are controlled by an FPGA, which provides their driving signals. For two of the EO switches (OS2 and OS5 in Fig. 2 of the main text), these signals are further modulated by outputs from RNGs. The clock output from the DSG (10 MHz) serves as the global synchronization signal for the entire system. The time delay of each electronic control loop can be measured with a precision of a few nanoseconds—well below the time scale defined by the manipulation rate of the time-bin qubit.

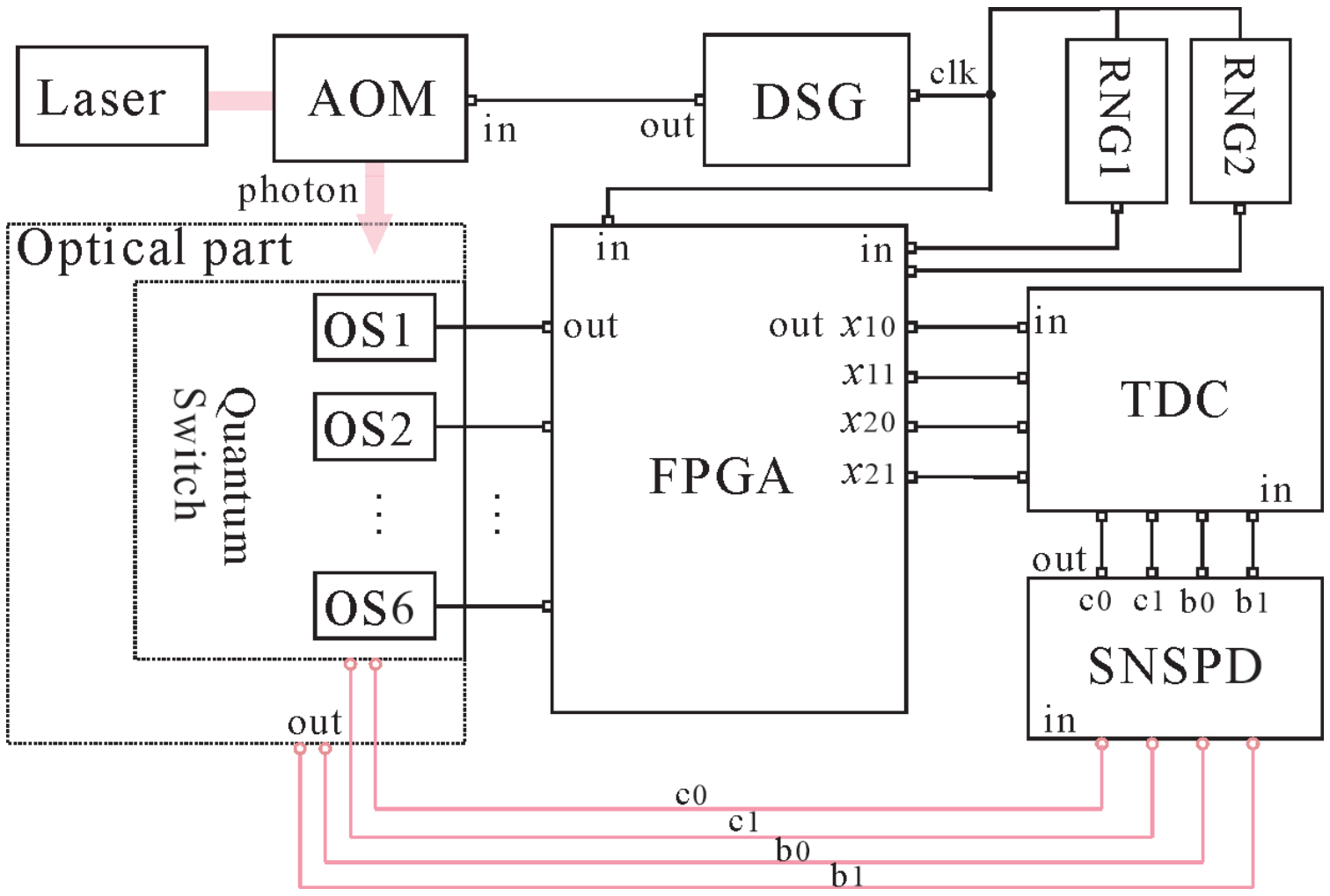

**Figure S2**: **Electronic control loop of our setup.** The AOM is driven by a DSG, whose clock signal is used to synchronize RNGs, FPGA, and OSs. RNGs are used to randomize the repreparation setting choices of Alice 1 and Alice 2. To read out Alices' measurement outcomes and repreparation setting choices, four pulse signals, labeled as $x_{ij}$ ($i \in \{1,2\}, j \in \{0,1\}$), are used to trigger the TDC. These trigger signals are synchronized to Alices' repreparation setting choices. DSG, digital signal generator; AOM, acoustic optical modulator; FPGA, field-programmable gate array; OS, electro-optical switch; RNG, random number generator.

Building on this synchronized architecture, we now analyze the detailed timing requirements for photon propagation through the quantum switch and the corresponding operations of the EO switches (OS1-OS6 in Fig. 2 of the main text). Here, OS1 and OS4 are located in the measurement AMZIs of the two Alices, OS2 and OS5 are placed in their repreparation AMZIs, and OS3 and OS6 guide the photons either within the quantum switch or out of it. Each experimental trial lasts 20 $\mu$s, during which only a single pair of entangled photons is generated. As discussed above, the generation of the pump pulse, the RNGs, and the measurement-and-repreparation operations performed by Alice 1 and Alice 2 are synchronized by a master oscillator (see Fig. S2). Each trial begins with the generation of a pump pulse at $t = 0$. When the control qubit is in the state $|H\rangle$, the signal photon reaches OS1 at $t = 5$ $\mu$s. After passing through the short or long arm of the first AMZI, it reaches OS2 at the earliest at $t = 5$ $\mu$s and at the latest at $t = 6$ $\mu$s. Subsequently, the photon reaches OS3 between $t = 5$ $\mu$s and $t = 8$ $\mu$s, OS4 between $t = 10$ $\mu$s and $t = 13$ $\mu$s, OS5 between $t = 10\mu$s and $t = 15$ $\mu$s, and OS6 between $t = 10$ $\mu$s and $t = 17$ $\mu$s. On the other hand, when the control qubit is in the state $|V\rangle$, the signal photon reaches OS4 at $t = 5$ $\mu$s. It then reaches OS5 between $t = 5$ $\mu$s and $t = 7$ $\mu$s, OS6 between $t = 5$ $\mu$s and $t = 9$ $\mu$s, OS1 between $t = 10$ $\mu$s and $t = 14$ $\mu$s, OS2 between $t = 10$ $\mu$s and $t = 15$ $\mu$s, and OS3 between $t = 10$ $\mu$s and $t = 17$ $\mu$s. The asymmetry between the two cases arises from the different delays in the measurement AMZIs: the relative delay in Alice 2's AMZI is 2 $\mu$s, while that in Alice 1's AMZI is 1 $\mu$s.

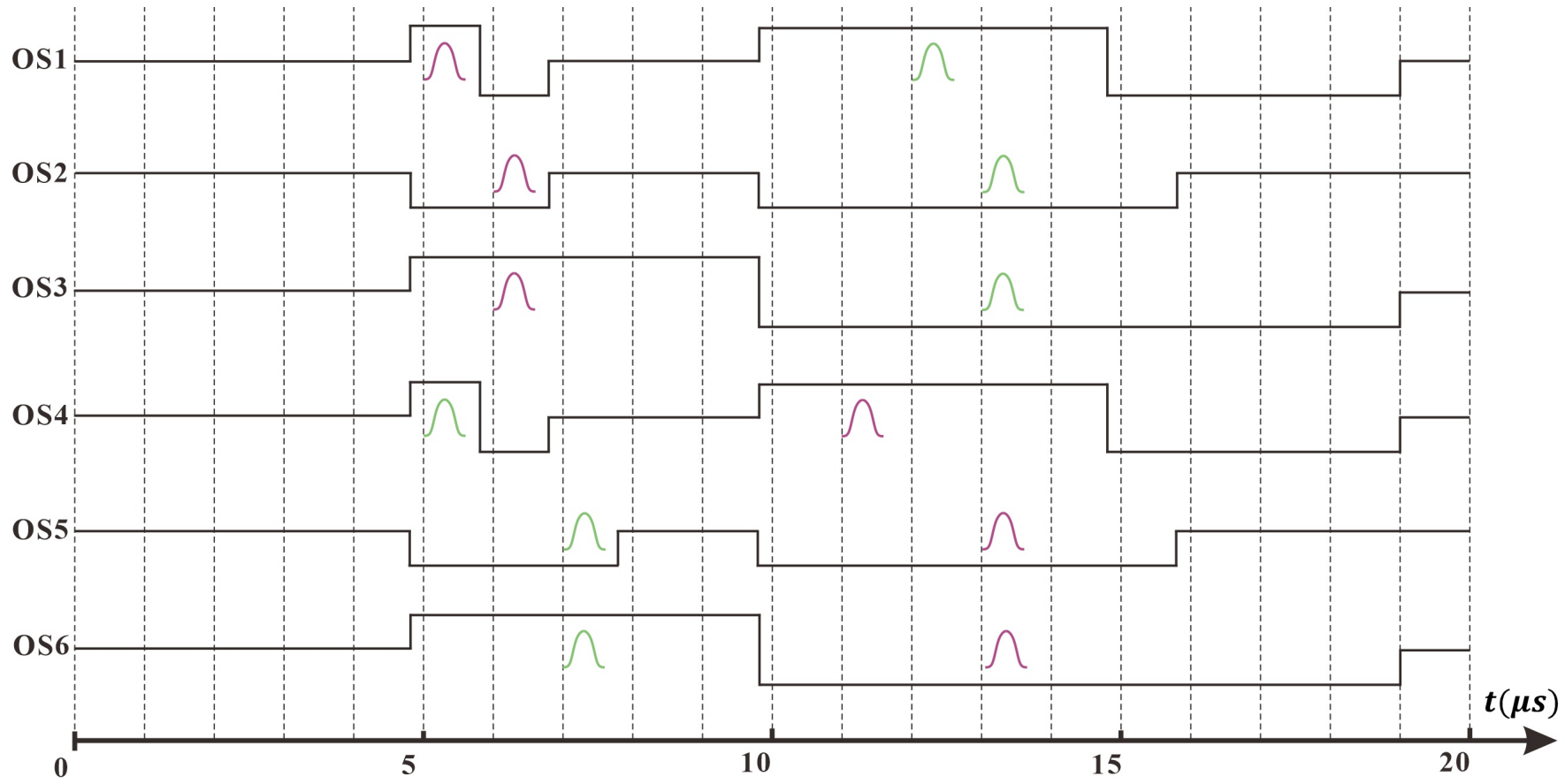

**Figure S3**: **Typical timing diagram of the driving signals for the EO switches when both Alices measure the initial target qubit $|0\rangle$ in the computational basis and reprepare it in the state $|0\rangle$.** For OS1, OS2, OS4, and OS5, a high-level (3.3 V) driving signal routes the photon to the long arm of the corresponding AMZI, while a low-level (0 V) signal routes it to the short arm. For OS3 and OS6, the high level routes the photon between Alice 1 and Alice 2, whereas the low level routes it out of the quantum switch. All optical switches are initialized at 1.7 V and reset to 1.7 V after $t = 19$ $\mu$s.

These considerations determine the timing requirements for setting the EO switches. Given that the pulse width is 600 ns, OS1 must be set before $t = 5$ $\mu$s and its driving signal must be maintained until $t = 14.6$ $\mu$s, while OS4 must be set before $t = 5$ $\mu$s and maintained until $t = 13.6$ $\mu$s. Similarly, OS2 and OS5 must be set before $t = 5$ $\mu$s and their driving signals must be maintained until $t = 15.6$ $\mu$s. The switches OS3 and OS6 must be triggered twice during each trial. The first activation routes the photon between Alice 1 and Alice 2 after the photon passes through the switches for the first time, while the second activation guides the photon out of the quantum switch after the second passage. In our experiment, OS3 is first set before $t = 5\ \mu$s and its driving signal is maintained until $t = 8.6\ \mu$s; it is then set again before $t = 10\ \mu$s and maintained until $t = 15.6\ \mu$s. Similarly, OS6 is first set before $t = 5\ \mu$s and maintained until $t = 9.6\ \mu$s, and is then set again before $t = 10\ \mu$s and maintained until $t = 15.6\ \mu$s. In Fig. S3, we present a representative timing diagram of the driving signals for the EO switches, corresponding to the case where Alice 1 and Alice 2 both measure the target qubit $|0\rangle$ in the computational basis and subsequently reprepare it in the state $|0\rangle$. In this diagram, all driving signals are initially set to 1.7 V and are reset to 1.7 V after $t = 19\ \mu$s. The rising time of all EO switches is less than 100 ns. The pulses indicate the arrival times of the photon at the corresponding EO switches: the pink pulses correspond to the order from Alice 1 to Alice 2, while the green pulses correspond to the order from Alice 2 to Alice 1.

The above timing analysis specifies the coordination required for the experimental operations. We now discuss the causal relations implied by this configuration. In the present experiment, true spacelike separation between Bob and the other three agents is not achieved, due to the lack of experimental equipment capable of performing rapid polarization measurements (*55, 56*). With such equipment, the current experimental configuration could be directly extended to realize genuine spacelike separation. Potential challenges in such an implementation may include phase disturbances, polarization fluctuations, and precise time synchronization among the devices, which are required to coordinate the operations of the agents in a field experiment.

# 5 The no-signaling restriction induced by the causal structure of our proposal.

The DRF inequality in Ref. (*53*) is built in the scenario with four agents involved, where two are embedded in a quantum switch Alice 1 and Alice 2, and the other two are space-like separated (Bob) or in the causal future (Charlie) from the quantum switch. In addition to the violation of the DRF inequality, experimental evidence of indefinite causal order requires that the correlations observed in the experiment obey some no-signaling conditions induced by the causal structure of the four agents. The no-signaling restriction requires that the mean value of the outcomes of an observer remain constant when other observers change their measurement settings. The requirement of the overall causal structure on the no-signaling condition can be expressed as

$$\mathcal{NS} := \{p \in \mathcal{P}_{\vec{a}bc|\vec{x}yz} : \vec{a}\,c \perp\!\!\!\perp_p y \;\&\; b \perp\!\!\!\perp_p \vec{x}\,z\}\,, \tag{S1}$$

Where $\mathcal{P}_{\vec{a}bc|\vec{x}yz}$ is the set of conditional probability distributions and $\perp\!\!\!\perp$ denotes statistical independence. For the DRF inequality we tested (Eq. (1) of the main text), the first term of it considers the situation where Bob projects his shared photon onto the eigenstate $|H\rangle$ of the Z basis ($y = 0$, $b = 0$). This reduces the switch to a wiring in which Alice 1 is before Alice 2. The no-signaling condition for this term is

$$\mathcal{DRF}_1 := \{p \in \mathcal{NS} : a_1 b \perp\!\!\!\perp_p x_2 \;\&\; \vec{a}\,b \perp\!\!\!\perp_p z\}\,. \tag{S2}$$

Analogously, the second term indicates a wiring from Alice 2 to Alice 1 and the no-signaling condition is

$$\mathcal{DRF}_2 := \{p \in \mathcal{NS} : a_2 b \perp\!\!\!\perp_p x_1 \;\&\; \vec{a}\,b \perp\!\!\!\perp_p z\}\,. \tag{S3}$$

# 6 Extending to test other DRF inequalities.

In addition to the VBC inequality tested in our experiment, Ref. (*53*) also introduced several other DRF inequalities. Some of these can be (weakly) violated by the quantum switch, while others cannot. All of these inequalities are valid and tight for the DRF polytope, which characterizes correlations compatible with a hidden-variable model satisfying 'Definite Causal Order', 'Relativistic Causality', and 'Free Interventions'. These inequalities define faces—though not necessarily facets—of a lower-dimensional projection of the full DRF polytope. Our experimental setup can be naturally extended to test any of the inequalities that are violated by the quantum switch. This is because the correlations involved in these inequalities are fully determined by the joint probabilities $P(a_1, a_2, b, c|x_1, x_2, y, z)$, all of which are accessible in our setup. The only modification required is the implementation of different single-qubit measurements on the control qubit, depending on the specific inequality being tested.